\documentclass[twocolumn,showpacs,preprintnumbers,amsmath,amssymb]{revtex4}

\usepackage{graphicx}
\usepackage{bm}

\begin{document}

\title{Solution of the Unanimity Rule on exponential, uniform and scalefree networks: \\
A simple model for biodiversity collapse in foodwebs}

\author{Rudolf Hanel$^{1}$, Stefan Thurner$^{1,2}$ \email{thurner@univie.ac.at}}
\email{thurner@univie.ac.at}

\affiliation{
$^1$ Complex Systems Research Group; HNO; Medical University of Vienna; 
W\"ahringer G\"urtel 18-20; A-1090; Austria \\
$^2$ Santa Fe Institute; 1399 Hyde Park Road; Santa Fe; NM 87501; USA 
} 

\begin{abstract}
We solve the Unanimity Rule on networks with exponential, uniform and scalefree  degree distributions. 
In particular we arrive at equations relating the asymptotic number of nodes in one of two states 
to the initial fraction of nodes in this state. The solutions for exponential and uniform networks  are exact, 
the approximation for the scalefree case is in perfect agreement with simulation results. We use these 
solutions to provide a theoretical understanding for experimental data on biodiversity loss in foodwebs, which 
is available for the three network types discussed.  The model allows in principle to estimate the critical value of 
species that have to be removed from the system to induce its complete collapse.

\end{abstract}


\maketitle

\section{Introduction}
Unanimity rule (UR) is generally associated with models, where a (binary) state of an 
atom or agent can change only  if all of its direct neighbors are in the other state, respectively. 
Usually UR is formulated in a network framework, where a node becomes 'activated' 
{\em only if} all the nodes pointing to it -- through directed links -- are 'activated'.  
These models have attracted some recent interest because of a number of 
important real world applications. 
Maybe the most relevant example is the modeling of biodiversity based on foodwebs. 
Species are nodes in this web. If one species is food for another species this is indicated 
by a directed link in the foodweb, pointing from the eaten species to the eater.  
Usually one species does not depend on a single other species but in general 
has a more diversified menue. In the picture of the foodweb this means that 
each node is pointed at from several other neighbors.
Imagine that all species in a hypothetical foodweb exist in one 
of two states: alive or extinct. If all the neighbors pointing to a particular node $i$, are 
extinct the node itself has no more food to live on and will go extinct in the next timestep 
as well; the relation to the UR becomes apparent. UR has been studied experimentally 
in actual foodwebs to model biodiversity loss \cite{dunne1,dunne2}. 
Earlier efforts on modeling  foodweb topology and its fragility 
have been conducted with respect to clustering, fragmentation, robustness, and degree distribution 
under the assumptions of random and non-random extinction of species \cite{SoleMontoya_2001}.
In \cite{MontoyaSole_2002} small world effects are investigated and 
it was concluded that foodwebs are not random networks with Poissonian degree distribution. 
Niche models -- as a method of sampling surrogate foodwebs -- are studied in 
\cite{Martinez2000,CamachoGuimera_2002_PRL,StoufferCamacho_2006} 
which (in the low connectancy limit)
display robust scaling properties of foodwebs which is in good
agreement with data from several field studies, see e.g. \cite{field}.
Further, their conclusion that the degree distribution of foodwebs decays exponentially rather than 
scale-free is in good agreement with \cite{dunne1,dunne2}.
Studies concerned with the robustness of foodwebs generally use the UR as an 
update mechanism, propagating initial extinction of species.

Surprisingly, the dynamics of  Unanimity Rule, which is   
a generalization of the Majority Rule of opinion dynamics \cite{redner,red,lambi,lambi1,lambi2} 
and reminds on features of the Voter model \cite{voter0,voter1,voter2,voter3}, the 
Axelrod model \cite{axelrod,axelrod2} as well as of 
Boolean networks \cite{kauffman,kauz}, 
is poorly known \cite{hanel,hanel2} and has been put on a more mathematical
basis only recently \cite{lambiotte07}. The unanimity model as presented there can also 
be viewed as a limiting case of a model for decision making scenarios \cite{watts0}. 
Note that UR differs from these previous models by the fact that it is {\em irreversible}, 
i.e. once a node has reached the activated state, it remains in it.  
This irreversibility of UR makes it an excellent candidate not only for modeling biodiversity as above 
but also for the adoption of  new technologies, such as MMS \cite{mms}, by interacting customers. 
Technological standards are generally irreversible once they are adopted by a population. 
Another specificity of UR is the fact that it is purely deterministic, i.e. once the topology 
of the underlying network is  fixed and an initial number of nodes are activated, 
the entire dynamics is determined.  In contrast, the Voter model, when applied to   
complex networks, involves a random step when a node chooses an interaction partner 
among its neighbors. Similarly, in the Majority Rule, a node choses randomly  between two nodes  
among its neighbors to form a majority triplet. 

In this paper we start from the dynamical equation for UR on networks and solve it analytically 
for the special cases of exponenential and
scale free networks. We then compare the results with experimental findings 
obtained from actual foodwebs \cite{dunne1,dunne2}. 

\begin{figure*}
 \includegraphics[width=2.0in]{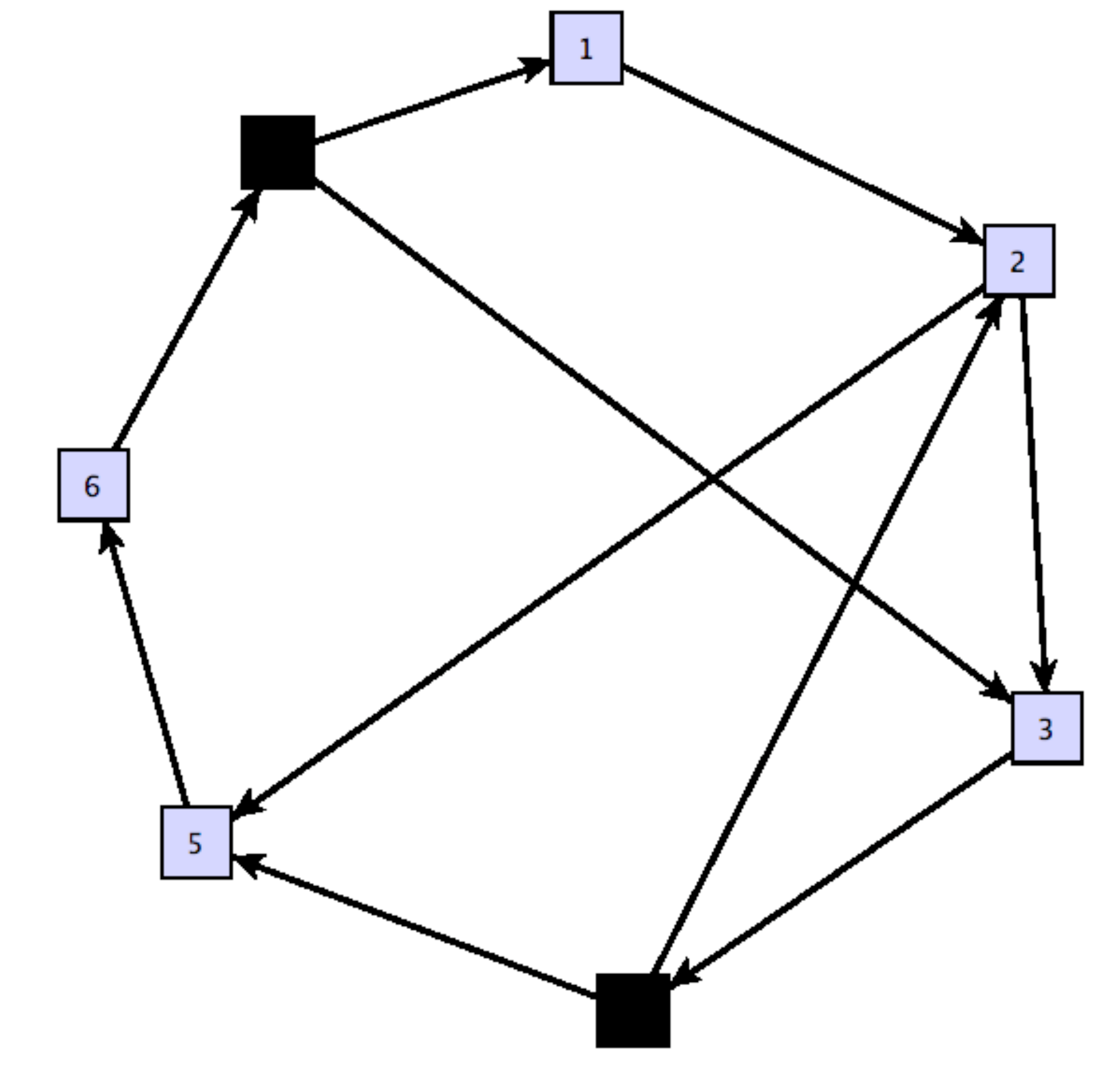}
 \hspace{0.6cm}
 \includegraphics[width=2.0in]{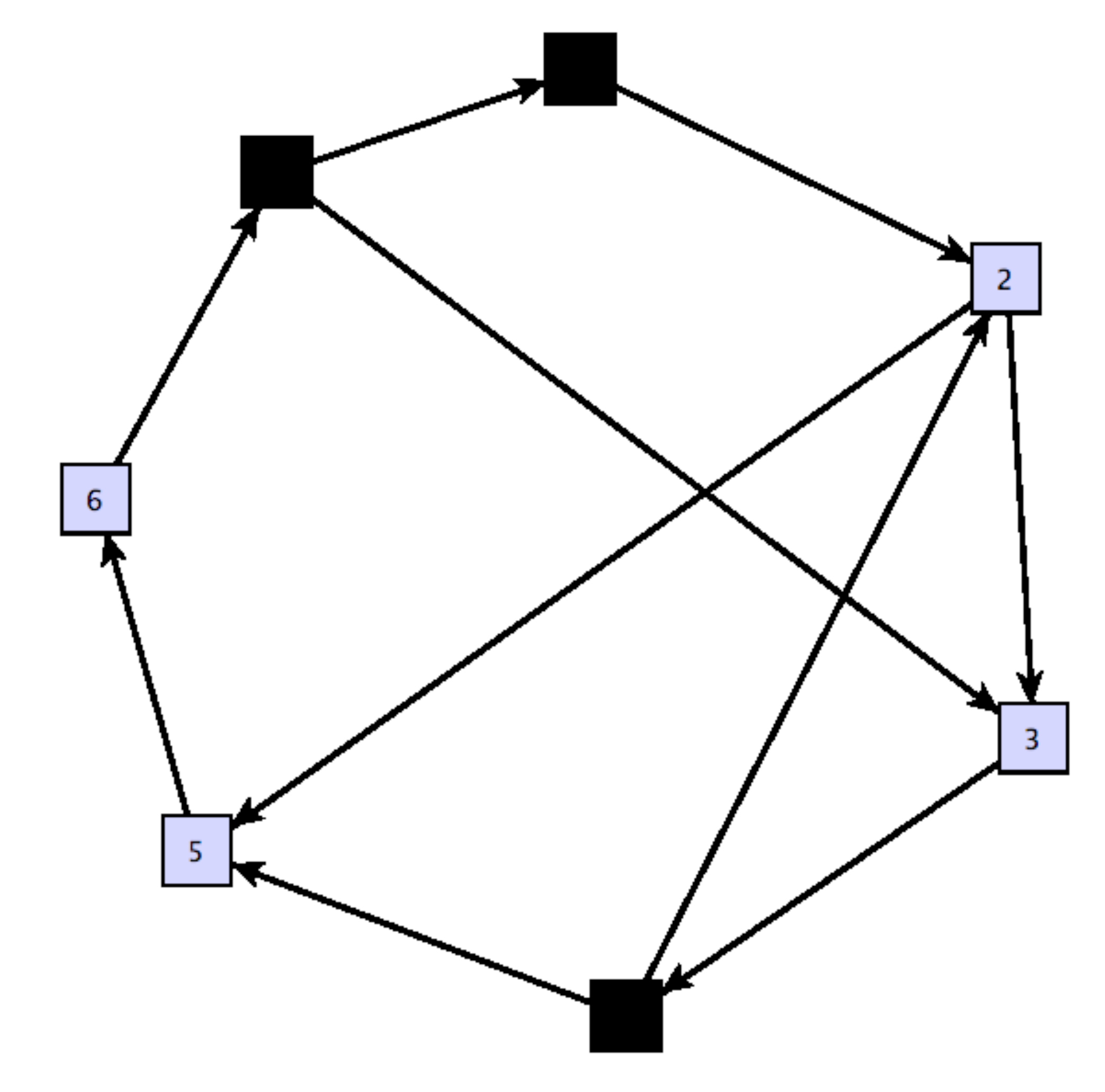}
 \hspace{0.6cm}
 \includegraphics[width=2.0in]{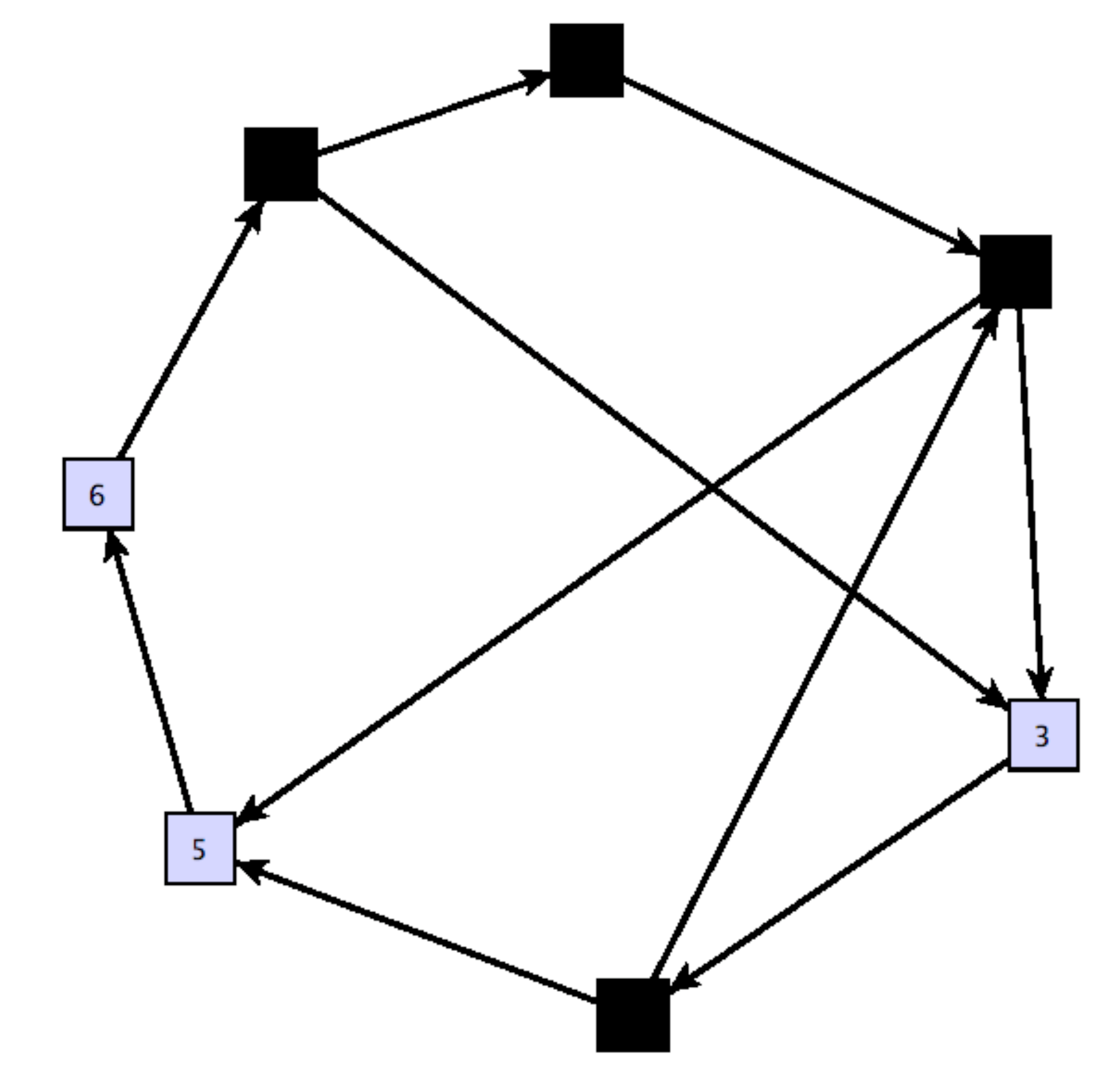}
 \caption{  
 First two steps of UR starting from an initial network of 7 nodes, 2 of them being activated. Initially 
 there is only one node among the non-activated nodes that satisfies the unanimity rule. It gets therefore 
 activated at the first time step. At that time, there is a new node whose 2 incoming links come from activated nodes. 
 It gets  activated at the second time step. This system gets fully activated at 
 the fourth time step. In the context of foodwebs for example node 2 eats species 1 and 4, 
 and serves as food for species 3 and 5. A black node means that it is extinct. (From \cite{lambiotte07}).}
 \label{fig1}      
\end{figure*}

\section{Unanimity Rule}

UR is implemented amongst agents in a network composed of $N$ nodes 
connected through $L$ directed links. Each node exists in one of two states: 
activated or inactivated.  
The  number of nodes with indegree $i$ (number of links pointing to it) 
is denoted by $N_i$ and depends on the underlying network structure, 
i.e. the indegree distribution is given by $p_i=N_i/N$.  
It is a fixed quantity that does not evolve with time.
The fraction of activated nodes at time $t$ we denote by $a_t$, 
for the fraction of activated nodes  with an indegree $i$, we write $a_{i,t}$. 
Initially (at $t=0$) there is a fraction of $a_0$ activated nodes. 
Obviously
 \begin{equation}
   a_t =\sum_{i} p_i a_{i,t}
 \end{equation}
holds.
The unanimity rule is defined as shown in  Fig. 1. 
If all the links arriving to an  inactivated node $i$ originate from  nodes which are activated at $t-1$, 
$i$ gets activated at $t$. Otherwise, it remains inactivated. 
At each time step every node is considered for an update, the dynamics is synchronous.
The process is iterated until the system reaches a stationary, {\em frozen} state, characterized by an 
asymptotic value $a_{\rm final}\equiv a_{\infty}$. 
The UR problem is to predict $a_{\infty}$ from the knowledge of the $a_0$ and the structure 
(indegree distribution) of the network. 

To solve problems of this nature a two step approach was suggested \cite{hanel,hanel2}: first map the 
problem onto an update equation, second find the asymptotic solutions of the latter. 
To derive the update equation, note that the probability that $i$ randomly chosen 
nodes are initially activated, is $a^i_0$ ($i$ is an exponent).  The average number of 
nodes with indegree $i$ and which respect the unanimity rule is therefore $N_i a^i_0$. 
Among these nodes,  $N_i a_0 a^i_0$ were already activated initially. This is due to the 
fact that  the total number of nodes with (in)degree $i$ which are initially activated is $N_i a_0$. 
Consequently, the number of  nodes  that gets 
activated at the first time step is
 \begin{equation}
 \label{delta0} 
   \Delta_{i,0} = \left(N_i - N_i a_0\right) a^i_0 \quad   , 
 \end{equation}
and, on average, the total number of occupied nodes with indegree $i$ evolves as
 \begin{equation}
    A_{i,1} = A_{i,0} + \Delta_{i,0} \quad.
 \end{equation}
 At the next time step, the average number 
of nodes with indegree $i$, which respect the unanimity rule and which are outside the initial set is 
$(N_i- N_i a_0) a^i_1$. Among those nodes, $\Delta_{i,0}$ have already been activated during the first 
time step, so that the average number of nodes which get activated at the second time step is
 \begin{equation} 
  \label{delta1}
   \Delta_{i,1} = (N_i - N_i a_0) (a^i_1 - a^i_0)  \quad .
 \end{equation}
Note that Eq. (\ref{delta1}) is valid because no node in $\Delta_{i,1}$ also belongs to $\Delta_{i,0}$. 
This is due to the fact that each node can only be activated by {\em one} combination of $i$ 
nodes in our model, so that no overlap is possible between $\Delta_{i,1}$ and $\Delta_{i,0}$. 
By iterating it is straightforward to show that the contributions $\Delta_{i,t}$ read
 \begin{equation}
  \label{deltai} 
   \Delta_{i,t} = (N_i - N_i a_0) (a^i_t - a^i_{t-1}) \quad ,
 \end{equation}
with $a_{-1}=0$, by convention. The number of activated nodes evolve as
$A_{i,t+1} = A_{i,t} + \Delta_{i,t}$, and 
by dividing by $N_i$, one gets the equations for the fraction of 
activated nodes $a_i \in [0,1]$
\begin{equation}
\label{eqA}
 a_{i,t+1} = a_{i,t} + (1-a_0) (a^i_t - a^i_{t-1} ) = a_{i,0} + (1-a_0)a_t^i \quad.
\end{equation}
Now $a_t$ is the convex sum of Eq. (\ref{eqA})  with weights according to the indegree 
distribution $p_i$, i.e.  $a_t=\sum_i p_i a_{i,t}$.  Finally we get 
for the update equation
 \begin{equation}
 \label{eqB}
  a_{t+1} = a_0 + (1-a_0) \sum _i p_i  a^i_t  \quad .
 \end{equation}
Numerical solutions for $t\to\infty$ are found in Fig. \ref{fig2} for several different indegree distributions $p_i$
(symbols).

\section{Exponential and Scalefree}

Let us  focus on the special choices of  the exponential
($p_i \propto e^{-\lambda i}$),  and  scalefree  ($p_i \propto i^{-\lambda}$),   indegree distributions, 
to analytically uncover the behavior of Eq. (\ref{eqB}). 
The simpler cases, $p_i=\delta_{i1}$ and $p_i=\delta_{i2}$ have been solved in \cite{lambiotte07}.

For the {\em exponential} case we have to find an expression for the
term $\sum _{i=1}^{M} p_i  a^i_t$, where $M$ is the maximum indegree, i.e. $M\leq N$. 
With the exact identity, $\sum _{i=1}^{M} c^i=c(1-c^M)/(1-c)$,  the exact asymptotic equation for the
exponential case is found to be 
 \begin{equation}
 \label{eqExp}
  a_{\infty} = a_0 + (1-a_0) a_{\infty} \left(\frac{1-e^{-\lambda}}{1-e^{-\lambda M}}
  \frac{1-\left(e^{-\lambda}a_{\infty}\right)^M}{1-e^{-\lambda}a_{\infty}} \right)  \quad ,
 \end{equation}
where we first have set 
$c=\exp(-\lambda)a_{\infty}$ for the non-normalized contribution of $p_i a_t^i$, 
and second
$c=\exp(-\lambda)$, to account for the norm of the distribution $p_i$.
Finally we have taken the limit $t\to\infty$. 

We get two important limiting cases: 
First, the {\em uniform distribution} is recovered as the limit $\lambda\to 0$,  
 \begin{equation}
 \label{eqUniform}
  a_{\infty} = a_0 + (1-a_0) \frac{a_{\infty}}{M} \left(\frac{1-a_{\infty}^M}{1-a_{\infty}} \right)  \quad ,
 \end{equation}
where we have used the fact that $\lim_{\lambda\to 0}(1-\exp(-M\lambda))/(1-\exp(\lambda))=M$.

Second, the large system limit $M\to\infty$ is
 \begin{equation}
 \label{eqExpInfty}
  a_{\infty} = a_0 + (1-a_0) a_{\infty} \left(\frac{1-e^{-\lambda}}{1-e^{-\lambda}a_{\infty}} \right)  \quad .
 \end{equation}

\begin{figure}
 \includegraphics[width=8.5cm]{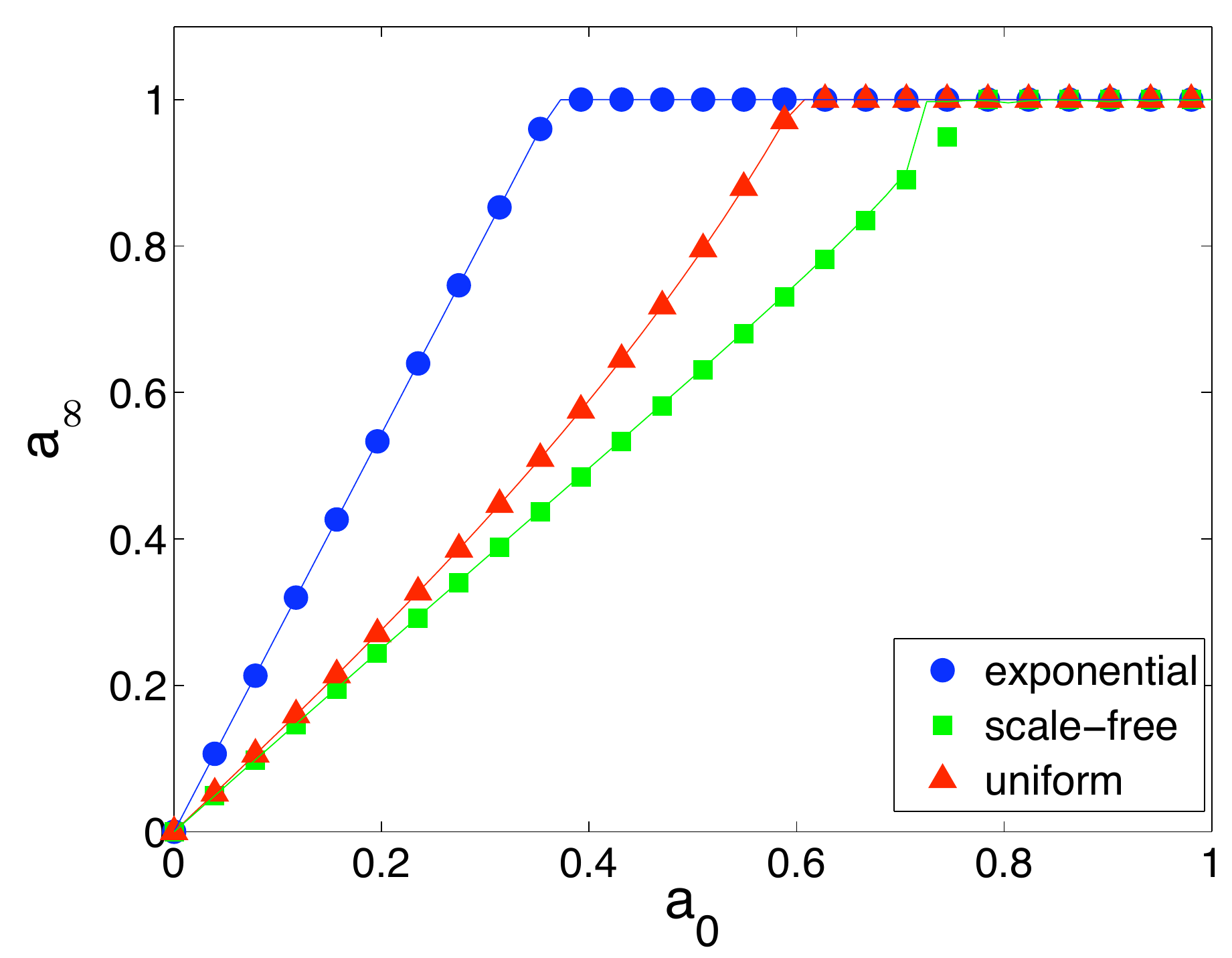}
 \caption{Asymptotic solutions, $a_{\infty}$, of the UR on three types of networks vs. initial condition $a_0$. 
 Symbols represent the numerical solutions of Eq. (\ref{eqB}). The networks used for this solutions have been 
 generated with $\lambda=-1$, $M=10$ for the exponential, $M=4$ for the uniform and  $\lambda=-0.5$, $M=10$ 
 for the scalefree case. Lines are the theoretical results of Eqs. (\ref{eqExp}), (\ref{eqUniform}) and (\ref{eqSF}), for the same 
 parameters. The critical value is where the plateau at $1$ is reached (all species extinct).
}
 \label{fig2}      
\end{figure}

The {\em scalefree} case is treated similarly. We approximate
$\sum_{i=1}^M i^{-\lambda}a_t^i\approx 1+\int_{3/2}^{M+1/2} dx\,\, x^{-\lambda}a_t^x$ and 
$\sum_{i=1}^M i^{-\lambda}\approx 1+\int_{3/2}^{M+1/2} dx\,\, x^{-\lambda}$. 
Defining the two-sided incomplete gamma-function as $\gamma(a,b,x)=\int_a^b dy\,\, y^{x-1} \exp(-y)$
the asymptotic solution for the scalefree unanimity rule reads
 \begin{equation}
 \label{eqSF}
  a_{\infty} = a_0 - (1-a_0)\bar\lambda\frac{
  1+\left|a_{\infty}\right|^{\bar\lambda} \gamma(\frac{3}{2}\left|a_{\infty}\right|,(M+\frac{1}{2})\left|a_{\infty}\right|, -\bar\lambda)}
  {\left(M+\frac{1}{2}\right)^{-\bar\lambda}-\left(\frac{3}{2}\right)^{-\bar\lambda}-\bar\lambda }  \quad ,
 \end{equation}
where $\bar\lambda=\lambda-1$. 
The quality of the results for the exponential, uniform and scalefree cases is seen in Fig. \ref{fig2}, 
where the asymptotic value is plotted against the initial $a_0$. Points represent the numerical 
solution to Eq. (\ref{eqB}) for  realizations of  networks of the characteristics specified in the figure caption.
Solid lines are the analytical results from Eqs. (\ref{eqExp}), (\ref{eqUniform}) and (\ref{eqSF}).
These equations show that the larger the exponents $\lambda$ (for exponential and scalefree)
the larger the critical value (where the plateau at $1$ is first reached) becomes. Similarly, the 
larger $M$ the larger the critical value.


\section{Experiment}

\begin{figure*}
 \includegraphics[width=2.0in]{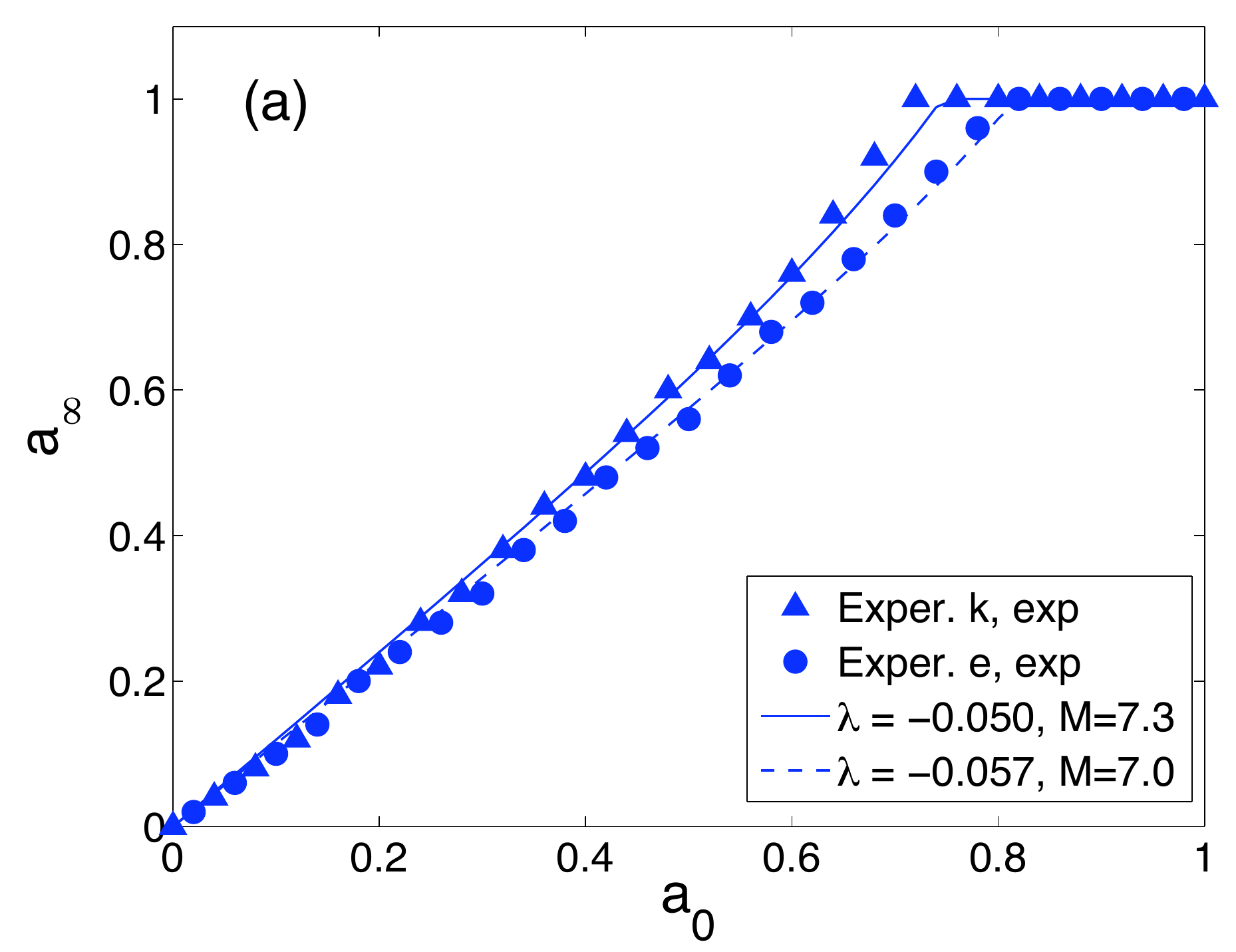}
 \includegraphics[width=2.0in]{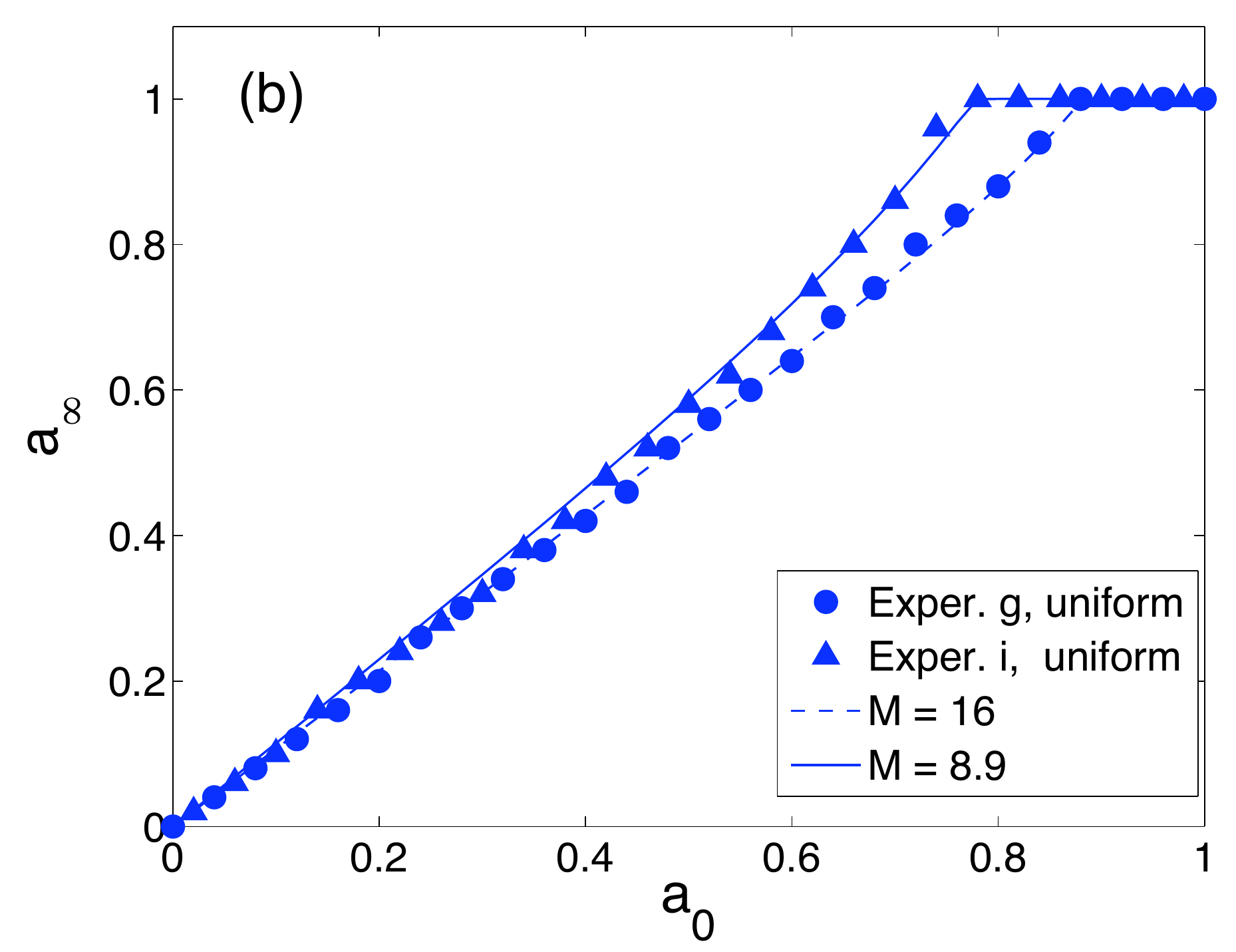}
 \includegraphics[width=2.0in]{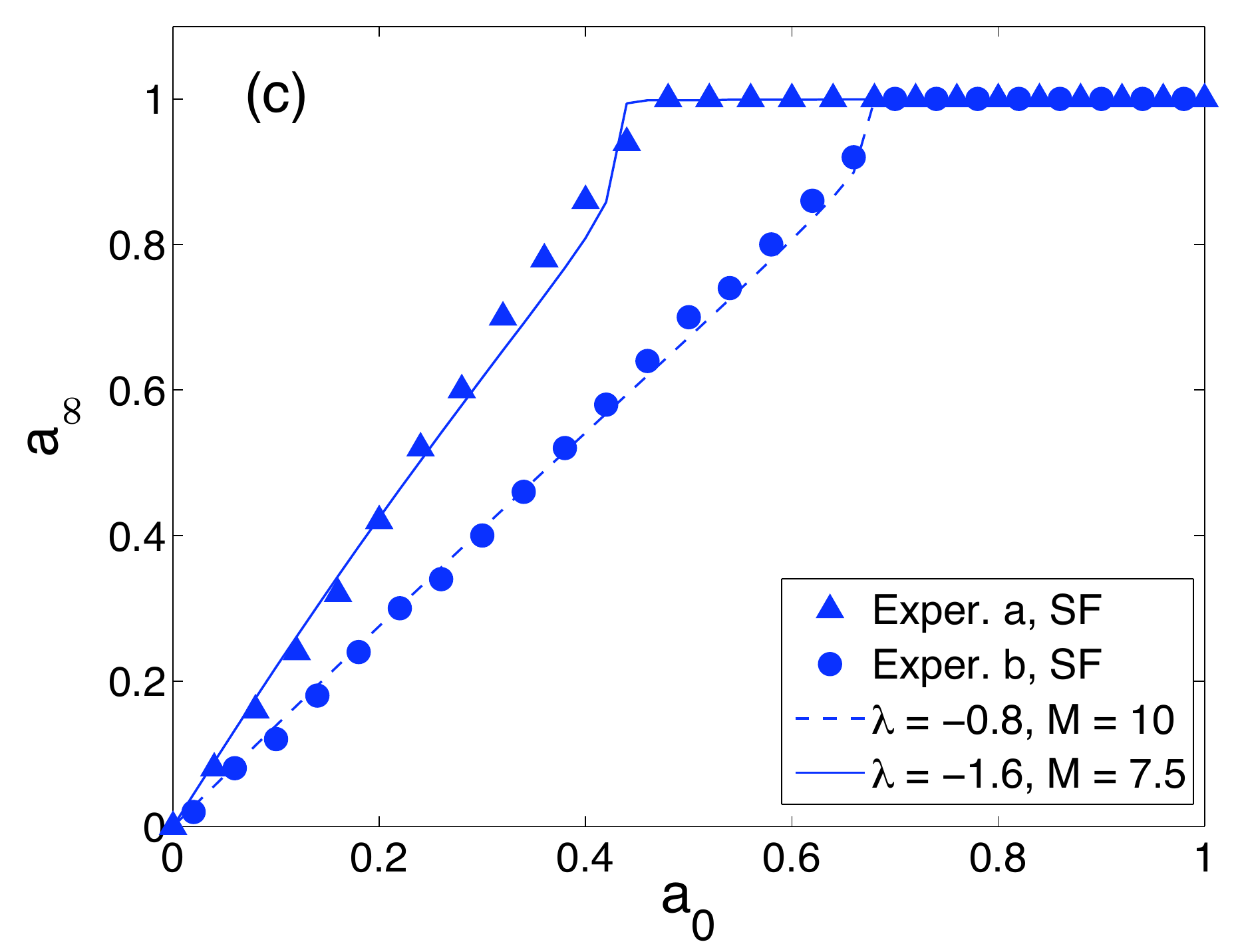}
 \caption{
Experimental data from \cite{dunne1,dunne2} (and references therein), 
for the number of asymptotically died out species 
on measured exponential (a), uniform (b) and scalefree (c) distributions (for randomly removed species). 
Lines are curves from our theoretical solutions with parameters as explained in the text. 
 }
 \label{fig3}      
\end{figure*}

In \cite{dunne1,dunne2} experimental results on foodwebs are presented, 
which are the outcomes from numerous field studies, see References 9,11,15,28-37 in  \cite{dunne2}. 
In \cite{dunne1} the number of species having died out as a result of the initial removal of 
a given fraction of the total population is presented. This constitutes a UR on networks, where -- in our 
notation -- the fraction of activated links $a_t$ are the ones that have  died out after $t$ 
timesteps. The reported 
networks there have been exponential, uniform and scalefree. We have re-drawn the data from 
\cite{dunne1} for two examples for each network type
\footnote{The choice for the particular experiments a, b, e, g, i, k comes from the 
fact that these ones  could be best extracted from the figures there.} 
in Fig. \ref{fig3} (symbols).  
Unfortunately, the network parameters can not get reconstructed from \cite{dunne2}. 
The data presented contains in- and outdegrees mixed together, which makes it impossible 
to estimate $\lambda$ and $M$ for the indegree only (which is needed for the UR).  The lines 
represent our theoretical results where we have estimated the parameters in the following way. 
For the exponential we fixed $\lambda$ to the values reconstructed from Fig. 2 in \cite{dunne2}, and varied $M$ 
until good fit was obtained. Of course $M_{\rm opt}$ is less than $M$ reported in  Fig. 2  of 
\cite{dunne2}.
The uniform case turned out to be optimal with $M_{\rm opt}$ about half of the reported value, 
the scalefree case was fitted after fixing $M$ to about half of the reported case and varying  $\lambda$. 

\section{Discussion}

We have solved the UR on exponential, uniform and scalefree networks. These 
solutions relate the structure of the underlying network together with an initial 
loss of diversity, $a_0$ with the final diversity in the system. Once the structure of the 
underlying network is known, these 
solutions allow further to find the critical value of initial 'species removal' $a_0^{\rm crit}$
at  which (and beyond which) a total collapse of the system will appear. 
The dynamics on foodwebs (who eats whom) is a particular example of this UR, 
the reported underlying networks are of  exponential, uniform and scalefree  nature. 
We have shown that not only do our results compare well with simulations but also 
with actual experimental data. The message of this work is to point out the potential 
danger of uncontrolled anthropogenic species removal.

{\bf Acknowledgements}
We are grateful to support from the Austrian Science Foundation projects P17621 and P19132.

\end{document}